\documentclass[floatfix,twocolumn,showpacs,preprintnumbers,amsmath,amssymb,superscriptaddress,prb,footinbib]{revtex4}
\usepackage{graphicx}% Include figure files
\usepackage{dcolumn}% Align table columns on decimal point
\usepackage{bm}% bold math
\begin{document}

\preprint{APS/123-QED}

\title{The electronic structure of La$_{1.48}$Nd$_{0.4}$Sr$_{0.12}$CuO$_4$ probed by high- and low-energy angle-resolved photoelectron spectroscopy: evolution  with probing depth}

\author{T. Claesson}
 \email{tcl@kth.se}
\author{M. M\aa{}nsson}
\author{A. \"{O}nsten }
\affiliation{
KTH Royal Institute of Technology, Materials Physics, Electrum 229, S-164 40 Kista, Sweden
}

\author{M. Shi}
\affiliation{Swiss Light Source, Paul Scherrer Institut,
CH-5232 Villigen PSI, Switzerland}

\author{S. Pailh\'{e}s}
\author{J. Chang}
\author{Y. Sassa }
\author{A. Bendounan}
\affiliation{Laboratory for Neutron Scattering, ETH Z\"{u}rich and
PSI Villigen, CH-5232 Villigen PSI, Switzerland}

\author{L. Patthey}
\affiliation{Swiss Light Source, Paul Scherrer Institut,
CH-5232 Villigen PSI, Switzerland}

\author{J. Mesot}
\affiliation{Laboratory for Neutron Scattering, ETH Z\"{u}rich and
PSI Villigen, CH-5232 Villigen PSI, Switzerland}

\author{T. Muro}
\author{T. Matsushita}
\author{T. Kinoshita}
\author{T. Nakamura}
\affiliation{ Japan Synchrotron Radiation Research Institute, SPring-8, Sayo, Hyogo 679-5198, Japan}

\author{N. Momono}
\author{M. Oda}
\author{M. Ido}
\affiliation{ Department of Physics, Hokkaido
University - Sapporo 060-0810, Japan}

\author{O. Tjernberg}
\affiliation{
KTH Royal Institute of Technology, Materials Physics, Electrum 229, S-164 40 Kista, Sweden
}

\date{\today}% It is always \today, today,

\begin{abstract}

We present angle-resolved photoelectron spectroscopy data probing the electronic structure of the Nd-substituted high-$T_c$ cuprate  La$_{1.48}$Nd$_{0.4}$Sr$_{0.12}$CuO$_4$ (Nd-LSCO). Data have been acquired at low and high photon energies, $h\nu$ = 55 and 500 eV, respectively. Earlier comparable low-energy studies of La$_{1.4-x}$Nd$_{0.6}$Sr$_{x}$CuO$_4$ ($x = 0.10, 0.12, 0.15$) have shown strongly suppressed photoemission intensity, or absence thereof, in large parts of the Brillouin zone. Contrary to these findings we observe spectral weight at all points along the entire Fermi surface contour at low and high photon energies. No signs of strong charge modulations are found. At high photon energy, the Fermi surface shows obvious differences in shape as compared to the low-energy results presented here and in similar studies. The observed difference in shape and the high bulk-sensitivity at this photon energy suggest intrinsic electronic structure differences between the surface and bulk regions.

\end{abstract}


\maketitle

\section{Introduction}

One of the most studied cuprate high-$T_{c}$ superconductor systems is the La$_2$CuO$_{4+\delta}$ system, which by the addition of Sr is hole-doped to give La$_{2-x}$Sr$_{x}$CuO$_4$ (LSCO). The possibility to grow high quality single crystalline samples over a very wide Sr doping range has enabled a large amount of experimental studies devoted to its electronic structure. Contributions have come from several diverse experimental fields, for instance neutron scattering \cite{christ07}, x-ray scattering \cite{zim98} and angle-resolved photoelectron spectroscopy (ARPES) \cite{zhou99}. Two of the investigated properties are the suppression of superconductivity in LSCO near $x=0.12$  and the possible existence of dynamic and static one-dimensional charge ordering (stripes) in LSCO and the Nd-substituted La$_{2-x-y}$Nd$_y$Sr$_{x}$CuO$_4$ (Nd-LSCO). ARPES has, much due to its ability to directly measure the spectral function $A(\vec{k},\omega)$, proven to be a very powerful tool in the studies of these and other central features of the electronic structure of the high-$T_{c}$ superconductors. 

The vast majority of ARPES studies on cuprate high-$T_{c}$ systems have been performed in the relatively low range of photon energies 20 - 100 eV. At these energies ARPES experiments suffer from a number of limitations and a certain amount of caution is necessary in the interpretation of these data. Particular attention should be paid to the limited probing depth at these low photon energies. This factor can pose a problem since a significant amount of the electron signal could come from other parts of the crystal structure than the deep lying Cu-O planes believed to be responsible for the mechanisms of high-$T_{c}$ superconductivity. This and other difficulties are to some extent overcome by carrying out ARPES measurements at high photon energies in the soft x-ray region. However, in this energy range other drawbacks appear, for instance the cross section for photoelectric effect is much lower and the obtainable energy and angular resolutions are significantly worse, as compared to the 20 - 100 eV photon energy range. Not until recently has the experimental progress permitted successful electronic structure studies using ARPES at soft x-ray photon energies \cite{Nccoprl, cu2oprb, fuji07}. Given the benefits and drawbacks of ARPES performed at low and soft x-ray photon energies, respectively, the combination of ARPES studies carried out in both energy ranges would provide a much more complete picture of the electronic structure in systems such as the strongly correlated cuprate superconductors. 

Here we report an ARPES study performed on the La$_{2-x-y}$Nd$_y$Sr$_{x}$CuO$_4$ ($x=0.12, y=0.4$) system at two different photon energies, $h\nu$ = 55 eV and $h\nu$ = 500 eV. The low photon energy dataset permits a detailed analysis of the near Fermi edge dispersion along the entire Fermi surface from the $d$-wave gap node direction to the antinode direction. On the other hand, the high photon energy data give access to the electronic structure at a much larger probing depth and enables a comparison to the low-energy data regarding such features as the shape and volume of the Fermi surface and the distribution of spectral intensity along it. 

\section{Experimental details} 

The single crystalline La$_{1.48}$Nd$_{0.4}$Sr$_{0.12}$CuO$_4$ (Nd-LSCO) samples, grown by the travelling solvent floating zone method, had a $T_c$ of 7 K. The high quality of the sample has been confirmed by neutron scattering measurements \cite{chang_ns07} performed on this particular batch. The samples were cleaved \textit{in situ} under ultra-high vacuum conditions (base pressure $<$ $2\times10^{-10}$ mbar) using a specially designed cleaver tool \cite{mans07} (low-energy data) and a glued on post (high-energy data), respectively. ARPES measurements were performed at two different synchrotron radiation facilities: (1) at the Swiss Light Source (SLS) synchrotron radiation facility using the Surface and Interface Spectroscopy beam line X09LA-HRPES \cite{fle04} and (2) at the SPring-8 synchrotron radiation facility using the BL25SU beam line \cite{sai00}. At the former beam line we have performed high-resolution spectroscopy ($\Delta$E = 50 meV) at $h\nu$ = 55 eV and T = 20 K using circularly polarized light. At the latter beam line we have performed high photon energy ($h\nu$ = 500 eV) ARPES measurements at an energy resolution of 100 meV at T = 20 K, also using circularly polarized light. Both ARPES end stations are equipped with Gammadata Scienta angle-resolving electron analyzers, the SLS end station has the SES-2002 model while the SPring-8 end station has the SES-200 model. The angular resolution along the analyzer slit was in both datasets on the order of $\pm0.1^{\circ}$, corresponding to momentum resolutions of 0.020 \AA$^{-1}$ and 0.006 \AA$^{-1}$. In the direction perpendicular to the analyzer slit, the momentum resolutions were on the order of 0.019 \AA$^{-1}$ and 0.030 \AA$^{-1}$ at low and high photon energies, respectively. 

\begin{figure} 
\includegraphics[width=\columnwidth]{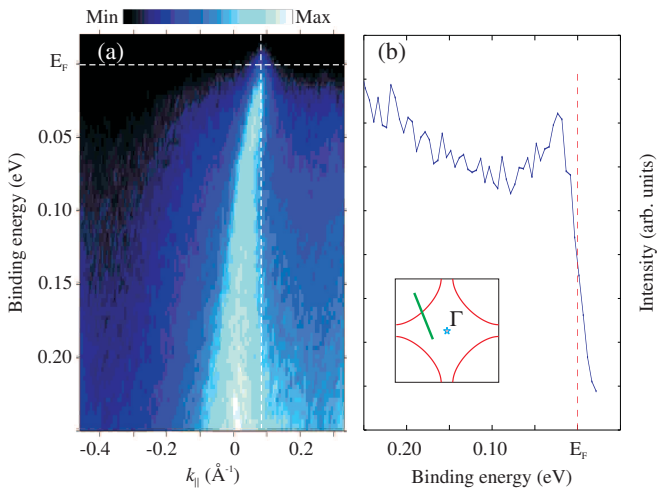} 
\caption{\label{fig1} (Color online) (a) ARPES spectrum acquired at $h\nu$ = 55 eV displayed as a color coded intensity map. Spectral intensity is displayed as a function of electron momentum and binding energy. The momentum-space orientation of this cut is displayed by the inset in (b). White dashed lines indicate the Fermi level and the momentum position of the energy distribution curve shown in (b). (b) Energy distribution curve acquired at the momentum indicated in (a). The solid red line in the inset is a schematic Fermi surface contour serving as a guide to the eye.} 
\end{figure}

\section{Results and discussion} 

Figure~\ref{fig1} displays ARPES data acquired at $h\nu$ = 55 eV. (a) presents a typical ARPES spectrum as a function of binding energy and electron momentum. A well-defined dispersive valence band feature crosses the Fermi level close to $k = 0.1$ \AA$^{-1}$. The momentum space orientation of this particular cut is close to the $d$-wave gap node direction, as displayed in the inset of (b). White dashed lines indicate the energy position of the Fermi level and the momentum location of the energy distribution curve (EDC) displayed in (b). The Fermi level has been determined by a Fermi function fit to photoemission data recorded from polycrystalline Cu on the sample holder. The EDC in the momentum position where the valence band crosses the Fermi level shows a pronounced peak structure with a clear Fermi cutoff. This experimental finding is contrary to that in a previous ARPES study on La$_{1.28}$Nd$_{0.6}$Sr$_{0.12}$CuO$_4$ (N.B. different Nd content as compared to our sample) performed by Zhou \textit{et al.} \cite{zhou99}. At the same level of Sr doping ($x$ = 0.12) these authors found no dispersing feature in the \mbox{$\Gamma$ - ($\pi, \pi$)} direction, where the cuprates usually show the most pronounced dispersion with well-defined quasiparticle peaks.

\begin{figure} 
\includegraphics[width=\columnwidth]{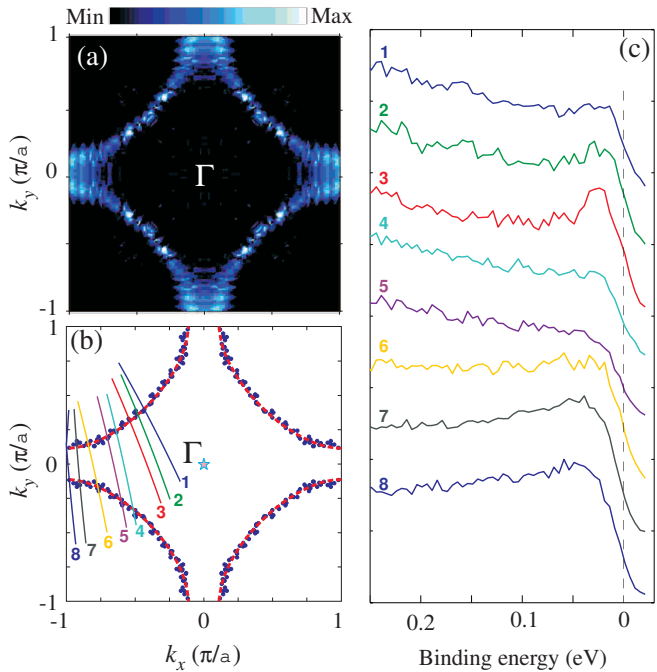} 
\caption{\label{fig2} (Color online) (a) Near Fermi level Nd-LSCO spectral intensity for the 55 eV data displayed as a color coded momentum space map. The valence band intensity was integrated over an energy window ranging from 10 meV below to 5 meV above the Fermi level. (b) Fermi level crossings of the valence band  as determined from functional fits of momentum distribution curves at $E_F$ to a Lorenztian function. The red dotted line indicates the Fermi surface resulting from a tight-binding expression for a CuO$_2$ plane. Tight-binding parameters are chosen so as to obtain a good fit to our data. (c) A series of angle-resolved energy distribution curves acquired at different positions along the Fermi surface contour. The shown EDC:s are acquired at the momenta where the corresponding numbered angular cut in (b) crosses the Fermi surface. } 
\end{figure}

Figure~\ref{fig2}(a) shows a color coded intensity map of the distribution of integrated spectral weight around the Fermi surface contour for the 55 eV data set. The valence band spectral intensity has been integrated over an energy window ranging from 10 meV below to 5 meV above the Fermi level. The acquired ARPES data have been folded into one quadrant of the two-dimensional Brillouin zone and symmetrized with respect to the \mbox{$\Gamma$ - ($\pi, \pi$)} line and finally repeated to cover the complete zone. Blue circular dots in Fig.~\ref{fig2}(b) display the Fermi surface at 55 eV determined from Lorentzian fits of momentum distribution curves at $E_F$. The red dotted line in (b) indicates a Fermi surface contour determined from the same tight-binding expression for the dispersion in a CuO$_2$ plaquette as in Ref.~\onlinecite{chang07}. The respective tight-binding parameter values are chosen to achieve a good fit with our experimental data \cite{tbnote}. Figure~\ref{fig2}(c) displays a series of angle-resolved EDC:s acquired at different momenta between the nodal and antinodal directions. The shown EDC:s are acquired at the momenta where the corresponding numbered angular cut in (b) crosses the Fermi surface.

As one moves along the Fermi surface from the nodal direction towards the ($\pi, 0$) antinodal point, following the numbered curves from 1 to 8 in Fig.~\ref{fig2}(c), the EDC:s show some evolution in terms of spectral shape. The near Fermi edge peak structure is most sharp in EDC number 3, slightly suppressed in 4 - 6 and finally starts to gain weight and gets more well-defined again close to the antinode in 7 and 8. These peaks are however much broader than those observed in La$_{2-x}$Sr$_{x}$CuO$_4$ ($x = 0.145, 0.17$) in the superconducting state \cite{chang07, shi07} and hence it is difficult to characterize them as quasiparticle peaks in a strict sense. Still it is possible to identify a more or less sharp peak in angle-resolved EDC:s for all momenta around the entire Fermi surface contour. This behavior contrasts to the results of Zhou \textit{et al.} \cite{zhou99}, who did not find any EDC:s with sharp peak structure along the entire Fermi surface contour, although some dispersion clearly was present in certain regions of the Brillouin zone.

The distribution of spectral weight integrated over a narrow energy window around the Fermi edge is shown in Fig.~\ref{fig2}(a). Before obtaining this intensity map, a linear background was subtracted from each momentum distribution curve in the ARPES spectra. The Fermi surface contour is centered around the $\Gamma$ point and has a volume reasonably consistent with the value $1 - x$ given by the Luttinger theorem \cite{lutt60}. With a filling level of 0.86 $\pm$ 0.05 the Luttinger theorem gives a doping level $x = 0.14$. Considering the limits given by experimental uncertainty, this value agrees well with the nominal doping level of our sample ($x = 0.12$). The shape of this Fermi surface contour and the distribution of intensity around it are both significantly different from the ARPES results of Zhou \textit{et al.} \cite{zhou99}, acquired at 55 eV photon energy from a La$_{1.28}$Nd$_{0.6}$Sr$_{0.12}$CuO$_4$ sample. Integration of their ARPES spectra over a 100 meV wide energy window around the Fermi level resulted in a concentration of intensity to straight patches located in a small momentum space region centered on the ($\pi$, 0) antinodal point. Our data also display a region of strong intensity around the ($\pi$, 0) point. However its shape is different from that in the data of Zhou \textit{et al}. In their data the boundary of spectral weight confinement is straight, while it is curved in our data. Near ($\pi$, 0) the locus of maximum intensity in Fig.~\ref{fig2}(a) closely follows the red dotted curve resulting from a tight-binding expression in (b). This behavior also contrasts to that in the data of Zhou \textit{et al.}, where the intensity reaches its maximum value in a small region of almost circular shape centered on the ($\pi$, 0) point.

Zhou \textit{et al.} \cite{zhou99} found only very weak spectral intensity in regions of the Brillouin zone other than the antinodal area. Especially in the vicinity of the ($\pi/2, \pi/2$) point the suppression of intensity was almost complete, giving a very poorly defined Fermi surface. A later ARPES study on La$_{1.4-x}$Nd$_{0.6}$Sr$_{x}$CuO$_4$ ($x = 0.10, 0.15$) performed by Zhou \textit{et al.} \cite{zhou01} indicated the presence of some spectral weight in this region, though strongly suppressed. However, our ARPES data when integrated over a 15 meV wide energy window result in evident spectral intensity along the entire Fermi surface, also in the vicinity of the ($\pi/2, \pi/2$) point. This is a fact one should note since it could give an indication on the strength of the charge modulations in Nd-LSCO. Strong suppression of spectral weight in the $d$-wave node direction, as observed in Ref.~\onlinecite{zhou99}, would point to a likewise strong modulation of charge. On the contrary, we observe strong spectral weight near the $d$-wave node location, which rather can be interpreted as the absence of strong charge modulations.

Furthermore, it is interesting to compare the shape of the near Fermi edge distribution of spectral weight from ARPES measurements with results from calculations. LDA calculations on the La$_{2-x}$Sr$_x$CuO$_4$ system \cite{xu87, sahra05} have indicated a notable valence band dispersion with respect to the out-of-plane component of electron momentum, $k_{z}$. Xu \textit{et al.} \cite{xu87} present such LDA results and show the Fermi surface topology for $k_z = 0$ and $k_z = \pi/c$. Our 55 eV ARPES results show a spectral weight distribution of reasonably similar shape as compared to the calculated results for $k_z = 0$. This is however not the case for the ARPES data of Zhou \textit{et al} \cite{zhou99}. The straight patches of intensity seen in their data are not reproduced by the calculation. Instead the LDA Fermi surface contour to a large extent runs more or less parallel to the [1,-1] direction, c.f. the red dotted curve in Fig.~\ref{fig2}(b) of the present study.

\begin{figure}
\includegraphics[width=\columnwidth]{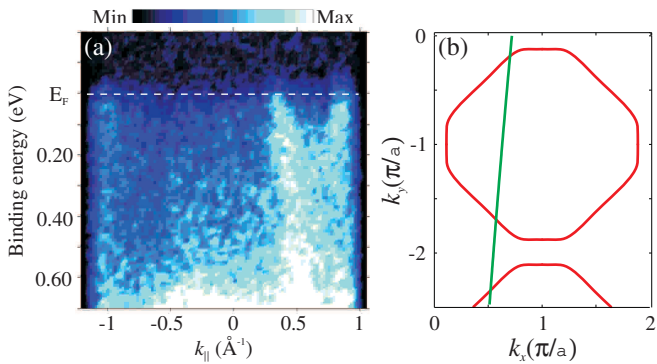} 
\caption{\label{fig3} (Color online) (a) ARPES spectrum acquired at  $h\nu$ = 500 eV displayed as a color coded intensity map. Photoemitted intensity is displayed as a function of binding energy and electron momentum. The dashed white line indicates the Fermi level. Two evident and one less apparent Fermi level crossings are shown. (b) Momentum space location of the angular cut (solid green line) displayed in (a). The solid red line is a schematic Fermi surface contour serving as a guide to the eye. } 
\end{figure}

Several aspects speak for the importance of highly bulk-sensitive photoemission measurements on the cuprate high-$T_{c}$ superconductors. Cleaving of an Nd-LSCO single crystal will always result in a sample with differences in electronic structure between the outermost surface layer(s) and the deeper lying bulk material. Such differences can definitely induce modifications in the electronic structure as probed by ARPES \cite{Nccoprl}. Moreover, the layered high-$T_c$ cuprate compounds have a large unit cell with one or several Cu-O planes, where the fundamental mechanisms of superconductivity are believed to originate. Virtually all ARPES measurements on high-$T_c$ systems have been interpreted as reflecting the electronic structure of these Cu-O planes. This assumption must however not necessarily be true. For instance, cleaving of a single crystal cuprate sample could easily result in a sample surface with the outermost Cu-O plane positioned at a depth of several \AA ngstr\"{o}ms beneath the cleaved surface. The commonly used range of photon energies for ARPES measurements on these systems is 20-100 eV, which makes the measurement more surface sensitive than bulk sensitive. In such a surface sensitive data set a considerable amount of the photoemission signal can originate in other crystal planes, located closer to the cleaved surface, than the Cu-O plane one is primarily interested in. 

In addition, there have been observations of a strong inelastic and photon energy dependent scattering in Sr and Ca core-level photoemission data from another compound in the cuprate family, namely Bi$_2$Sr$_2$CaCu$_2$O$_{8+\delta}$ \cite{zak03}. Scattering of this kind gives a reduced mean free path for the photoemitted electrons as they propagate from the point of photon absorption towards the sample surface. For emission from a deeply lying Cu-O plane this type of scattering could reduce the mean free path to such a degree that these electrons only would contribute a small part of the total photoemission signal. However, the influence such effects might have on photoemission data from the cuprate systems is largely reduced by performing experiments which make use of photon energies in the soft x-ray region. 

Moreover, in order to interpret the measured spectrum as the single-particle Green function one has to assume that the photoelectron upon excitation becomes effectively decoupled from the rest of the system. This assumption is readily justified at high energy but is more questionable at low energy.

Figure~\ref{fig3}(a) displays an ARPES spectrum from Nd-LSCO acquired at a photon energy of 500 eV. One can clearly discern two distinct dispersive valence band features crossing the Fermi level in the momentum region $k = 0.4$ to 1.0 \AA$^{-1}$. Weaker signs of another dispersive feature can also be noted near $k = -1.0$ \AA$^{-1}$. Even though the spectra of Figs.~\ref{fig1}(a) and \ref{fig3}(a) are collected using the same analyzer acceptance angles, a much larger part of momentum space is covered in the latter due to the use of the higher photon energy. The momentum space orientation of this particular cut is displayed in Fig.~\ref{fig3}(b). 

\begin{figure} 
\includegraphics[width=\columnwidth]{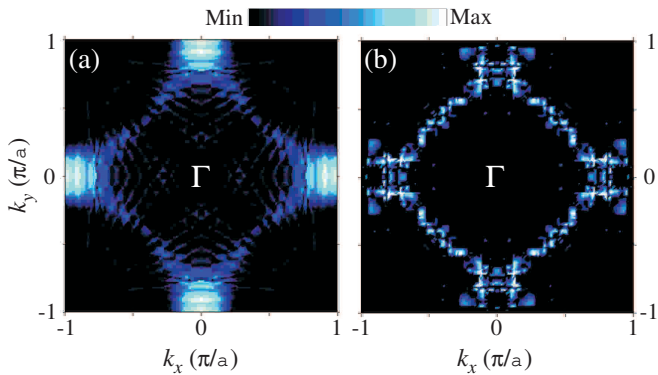} 
\caption{\label{fig4} (Color online) Color coded momentum space maps of near Fermi level spectral weight derived from ARPES data acquired at $h\nu$ = 55 eV (a) and 500 eV (b). In order to simulate the effect of a different energy resolution, the ARPES spectra on which (a) is based have been broadened by convolution with a Gaussian function.  The ARPES intensity is integrated over an energy window from 40 meV below to 30 meV above the Fermi level in both (a) and (b).  } 
\end{figure}

The combination of an inferior energy resolution and a lower count rate for photoemission experiments in the soft x-ray energy region makes it difficult to perform a detailed analysis in terms of EDC peak dispersion as was done for the low-energy data set. However, the near Fermi edge distribution of spectral weight is readily analyzed and can also be compared to the low-energy results. The two datasets (low and high energy) were obtained with energy resolutions of 50 meV and 97 meV, respectively. To be able to compare the two datasets, we have simulated the effect of a different energy resolution in the low-energy data in the following way. The low-energy ARPES spectra have been energy broadened by convolution with a Gaussian function whose full width at half maximum (83 meV) is calculated from the energy resolutions in the two datasets. Figure~\ref{fig4} displays color coded maps of spectral intensity integrated over an energy window ranging from 40 meV below to 30 meV above the Fermi level for the broadened low-energy data and the high-energy (500 eV) data. Before these intensity maps were obtained, a linear background was subtracted from each momentum distribution curve in the ARPES spectra. The spectral intensity displayed in this figure has been folded into the quadrant ($k_x, k_y > 0$) then symmetrized with respect to the $\Gamma$-($\pi,\pi$) line and finally duplicated to fill the entire Brillouin zone. 

As the Gaussian broadened intensity map of Fig.~\ref{fig4}(a) is compared to that based on the untreated ARPES spectra in Fig.~\ref{fig2}(a), one notices that the two figures are strikingly similar as far as both shape and distribution of spectral intensity are concerned. The only noticeable influence from the modified energy integration window and the energy broadening is a slight redistribution of intensity at the antinode. Hence, the use of different energy resolutions in the two current data sets is not affecting the Fermi surface properties studied here to any larger extent. Any differences such as the Fermi surface shape or the variation of spectral intensity must hence be attributed to other circumstances, such as the difference in photoemission probing depth. 

A more interesting comparison is that between the two intensity maps in Fig.~\ref{fig4}. At high energy (b), spectral weight is present along the entire contour from the nodal to the antinodal direction. This behavior is similar to our own 55 eV data but in contrast to the low-energy data presented in Ref.~\onlinecite{zhou99}. The high-energy contour encloses a region, whose volume is reasonably consistent with the Luttinger value $1 - x$. A filling level of 0.84 $\pm$ 0.05 corresponds to a doping value $x$ of 0.16 which, within the limits of experimental uncertainty, matches the nominal Sr doping level of our sample ($x = 0.12$). The energy broadened 55 eV data in the same way result in a filling level of 0.86 $\pm$ 0.04, corresponding to an actual doping level of 0.14, within the experimental uncertainty limits of the nominal doping value. The difference in measured doping level between 55 and 500 eV could indicate a difference in doping between the surface region and the bulk of the sample but it is difficult to draw such a conclusion in view of the experimental uncertainties involved.

Regarding the shape and distribution of spectral weight along the Fermi surface there are notable differences between the low- and high-energy data. In Fig.~\ref{fig4}(a) one notices the relatively large concentration of spectral intensity in a small area around the antinodal point, while the intensity is lower in other parts along the contour. The high-energy data of Fig.~\ref{fig4}(b) has a more uniform intensity distribution with the antinode region at about the same level as other parts of the contour. 

Turning to the shape of the Fermi surfaces, it is clear that there is a substantial difference between the more surface-sensitive data and the more bulk-sensitive data. The low-energy intensity in (a) approximately follows a contour whose centre of curvature is located at the ($\pi$,~$\pi$) point. In contrast, the large segment in (b) seems to follow a contour whose centre of curvature is positioned at the $\Gamma$ point. One also notices that the high-energy Fermi surface contour seems to be composed of two different segments. Besides the large segment already mentioned, there is also a small pocket centered at ($\pi$,~0). Experimental evidence for small Fermi surface pockets was recently observed in underdoped YBa$_2$Cu$_3$O$_{6.5}$ by means of Hall resistance oscillations \cite{doi07}. In light of this, one might speculate that the existence of such pockets is something common among underdoped cuprates.

The observed differences in the Fermi surface shape are clearly related to the difference in probing depth between low and high kinetic energies. This demonstrates that caution is necessary in interpreting low-energy photoemission data as indicative of bulk properties even in quasi two-dimensional systems such as the cuprates. As a matter of fact, comparable behavior has already been observed in related systems. A similar result with a modified Fermi surface shape was recently observed in soft x-ray ARPES data from Nd$_{2-x}$Ce$_x$CuO${_4}$ \cite{Nccoprl}. Highly bulk-sensitive hard x-ray photoelectron spectroscopy (XPS) data on La$_{2-x}$Sr$_x$CuO$_4$ and Nd$_{2-x}$Ce$_x$CuO${_4}$ \cite{tagu05b} have also indicated a modified electronic structure down to probing depths of several nanometers.

As was done for the 55 eV data, one can also compare the shape of the high-energy contour from our ARPES measurements with results from LDA calculations on the La$_{2-x}$Sr$_x$CuO$_4$ system \cite{xu87}. In these theoretical results, a notable dispersion of the Fermi surface contour with respect to the out-of-plane component of electron momentum, $k_z$, is observed. Based on this theoretical prediction it is hence possible to interpret the deviations in Fermi surface shape observed between the low- and high-energy data as partly due to probing of a different $k_z$ value. 

\section{Conclusions} 

To conclude, the present ARPES study of the Nd-LSCO system at low and high photon energies shows differences in the near Fermi edge spectral characteristics as compared to previous studies. In contrast to earlier results, we observe evident spectral weight along the entire Fermi surface contour. No regions with suppressed spectral intensity are found and hence nothing in our data supports the presence of strong charge modulations. However, based on this observation it is difficult to draw any distinct conclusions regarding the influence of stripes in this compound. In a similar ARPES study on La$_{1.28}$Nd$_{0.6}$Sr$_{0.12}$CuO$_4$ \cite{zhou99}, performed at $h\nu=$ 55 eV, virtually all spectral weight is confined to a small patch with linear boundaries located at the antinodal point. Contrary to this finding, the spectral intensity in our 55 eV data follows a curved, almost circular, contour from the nodal to the antinodal direction. 

Moreover, our ARPES data acquired at 500 eV photon energy show clear differences as compared to the low photon energy data. At high energy, the spectral weight seems to be more evenly distributed over the entire Fermi surface contour, not showing particularly elevated intensity close to the ($\pi$, 0) point as is observed in our 55 eV data. One can also note a distinct change in Fermi surface shape between the 55 and 500 eV data sets. In going from low to high photon energy, the centre of curvature shifts from ($\pi$,~$\pi$) to $\Gamma$ with a possible addition of pockets at ($\pi$,~0). The increase in photoelectron probing depth at soft x-ray photon energies suggests that this can be explained by intrinsic electronic structure differences between the bulk and surface regions. Similar behavior supporting this conclusion has already been experimentally observed in Nd$_{1.85}$Ce$_{0.15}$CuO$_4$ using soft x-ray ARPES \cite{Nccoprl} and hard x-ray XPS \cite{tagu05b}. 

% The ($\pi$, $\pi$) centered shape in the former case has at 500 eV turned into a $\Gamma$ centered shape with possible pockets at ($\pi$, 0). 

\begin{acknowledgments}

This research has been supported by the Swedish Research Council, the G\"oran Gustafsson Foundation, the Knut and Alice Wallenberg Foundation, the Foundation for Strategic Research (SSF), the Swiss National Science Foundation (through NCCR, MaNEP, and grant Nr 200020-105151), the European Union and the Japan Synchrotron Radiation Research Institute (JASRI) through a ``Budding Researchers Support Proposal'' (Proposal No. 2006B1020).

\end{acknowledgments}

% \bibliography{Ndlsco} % Refers to Coonio.bib
% Produces the bibliography via BibTeX.

\end{document}